\documentclass[12pt, a4paper]{article}
\usepackage{amsmath,amssymb,amsfonts}
\usepackage{graphicx}
\usepackage[colorlinks]{hyperref}
\usepackage{makecell}
\setcellgapes{3pt}

\begin{document}

\title{Induced vacuum energy density of quantum charged scalar matter in the background of an impenetrable magnetic tube with the Neumann boundary condition\\\phantom{hjbj}}%

\author{V.M. Gorkavenko${}^{1}$,  T.V. Gorkavenko${}^{1}$, Yu.A. Sitenko${}^{2,3}$, M.S. Tsarenkova${}^{1}$\\
\it \small ${}^{1}$Taras Shevchenko National University of Kyiv,
\\ \it \small 64 Volodymyrs'ka str., Kyiv 01601, Ukraine\\
\it \small ${}^{2}$Bogolyubov Institute for Theoretical Physics, National Academy of Sciences of Ukraine,
\\ \it \small 14-b Metrologichna str., Kyiv 03143, Ukraine\\
\it \small ${}^{3}$Donostia International Physics Center, 4 Paseo Manuel de Lardizabal,
\\ \it \small 20018 Donostia-San Sebastián, Gipuzkoa, Spain}

\date{}

\maketitle

\begin{abstract}
We consider vacuum polarization of charged scalar matter field outside the tube with magnetic flux inside. The tube is impenetrable for quantum matter and the perfectly rigid (Neumann) boundary condition is imposed at its surface. We write expressions for induced vacuum energy density for the case of  a space of arbitrary dimension and for an arbitrary value of the magnetic flux. We do the numerical computation for the case of half-integer flux value in  the London flux units and $(2+1)$-dimensional space-time.
We show that the induced vacuum energy of the charged scalar matter field is induced if the Compton wavelength of the matter field exceeds
the transverse size of the tube considerably. We show that vacuum energy is periodic in the value of the magnetic flux of the tube,
providing a quantum-field-theoretical manifestation of the
Aharonov-Bohm effect. The dependencies of the induced
vacuum energy upon the distance from the center of the tube under the different values of its  thickness  were obtained.
Obtained results are compared to the results obtained earlier in the case of the perfectly reflecting
(Dirichlet) boundary condition.
It is shown that  the value  of the induced vacuum energy density in the case of the Neumann boundary condition is  greater than in the case of the Dirichlet boundary condition.

\end{abstract}

\maketitle

\section{Introduction}

More than 70 years ago it was shown by Casimir \cite{Cas} that the presence of external boundaries leads to changes in the vacuum energy density.
First, two perfectly conducting plates at a very tiny distance apart were considered. It was shown that the difference between the vacuum expectation values leads to the emergence of a force of interaction between the plates. Since then, many setups of different boundaries' shapes and materials have been considered. The boundary manifolds are usually chosen as a disconnected noncompact (as the infinite plates), or, in other cases, a closed compact object (as a box or a sphere), see, e.g., \cite{Eli} -- \cite{Bordag1}. However, there is another case that is interesting of its own accord: a connected noncompact object (e.g., an infinite tube). 

As shown by  Aharonov and Bohm in the framework of first-quantized theory, see \cite{Aha}, magnetic flux inside an impenetrable for matter field cylindrical tube can interact with quantum matter outside the tube.
The consequences arising from it in the framework of the second-quantized theory are the polarization of the  vacuum and the induction of the vacuum current and magnetic flux outside the tube.
The effect of the boundary condition at the surface of the impenetrable tube and magnetic flux inside the tube on the vacuum of matter field outside the tube  has the name of the Casimir-Bohm-Aharonov effect \cite{Sit}.
The boundary condition in this setup affects the  matter field outside the tube essentially.

It should be noted the problem of vacuum polarization outside the impenetrable magnetic tube has numerous physical applications.
In astrophysics it can be considered as model of the cosmic strings, that may have appeared in the early Universe as a result of phase transitions with
spontaneous gauge symmetry breaking \cite{Kibble} -- \cite{Hindmarsh}. In condensed matter physics it can be considered as model of Abrikosov-Nielsen-Olesen vortex in superconductors of the second group, see, e.g., \cite{Abr,Nielsen} or as  disclinations in nanoconical structures, see, e.g., \cite{Krishnan} -- \cite{LowTemp}.

It should be noted that initially the Bohm–Aharonov effect was considered under the assumption that the transverse size of the tube is zero, which corresponds to the singular
magnetic vortex, see, e.g.,  \cite{Sit}, \cite{Ser} -- \cite{our2}.

In this paper, we  will consider the case of charged  scalar matter. In the case of finite transverse size, impenetrable magnetic  tube  boundary conditions can be generically parameterized 
with the use of a family of boundary conditions
of the Robin type
\begin{equation}\label{Robin}
    (\cos \theta\, \psi + \sin \theta\, r \partial_r \psi)|_{r_0} =0.
\end{equation}
Here cases $\theta=0$ and $\theta=\pi/2$ correspond to Dirichlet and Neumann boundary conditions respectively. 
For the induced vacuum energy, the case of the Dirichlet boundary condition  was considered in  \cite{our3} -- \cite{our2013}. For the induced vacuum current and magnetic flux,  the case of the Dirichlet boundary condition  was considered in \cite{our2016},  the case of the Neumann boundary condition  was considered in \cite{our2022} and the general case for the arbitrary value of the parameter $\theta$ was considered in \cite{our2022prd}.

In this paper, we will focus on the  vacuum polarization of the charged scalar matter outside the impenetrable finite-thickness magnetic tube   with  the Neumann boundary condition at its surface.

The paper is organized as follows. In the second section, we provide a general definition of the induced renormalized vacuum energy density for the quantized charged scalar field for the case of $d$-dimensional space. In the third section, using numerical methods, we compute the value of the induced vacuum energy density for the simplest case of $(2+1)$-dimensional space-time, namely outside  the impenetrable tube (it is a ring in 2-dimensional space) of radius $r_0$  and magnetic flux inside it. In the fourth section, we summarize and discuss the results.

\section{Energy density}

Lagrangian for a complex scalar field $\psi$  in
$(d+1)$-dimensional space-time has form
\begin{equation}\label{0}
\mathcal{L}=({\mbox{ $\nabla$}}_\mu\psi)^*({\mbox{
$\nabla$}}^\mu\psi)-m^2\psi^*\psi,
\end{equation}
where ${\mbox{ $\nabla$}}_\mu$ is the covariant derivative and $m$
is the mass of the scalar field.
The operator of the quantized charged scalar field is represented in
the form
\begin{equation}\label{a11}
\Psi(x^0,\textbf{x})=\sum\hspace{-1.4em}\int\limits_{\lambda}\frac1{\sqrt{2E_{\lambda}}}
\left[e^{-iE_{\lambda}x^0}\psi_{\lambda}(\textbf{x})\,a_{\lambda}+
  e^{iE_{\lambda}x^0} \psi_\lambda^\ast(\textbf{x})\,b^\dag_{\lambda}\right].
\end{equation}
Here, $a^\dag_\lambda$ and $a_\lambda$ ($b^\dag_\lambda$ and
$b_\lambda$) are the scalar particle (antiparticle) creation and
annihilation operators satisfying commutation relation; $\lambda$ is
the set of parameters (quantum numbers) specifying the state;
  $E_\lambda=E_{-\lambda}>0$ is the energy of the state; symbol
  $\sum\hspace{-1em}\int\limits_\lambda$ denotes summation over discrete and
  integration (with a certain measure) over continuous values of
  $\lambda$; wave functions $\psi_\lambda(\textbf{x})$ are the
  solutions to the stationary equation of motion,
\begin{equation}\label{a12}
 \left\{-{\mbox{\boldmath $\nabla$}}^2  + m^2\right\}  \psi_\lambda(\textbf{x})=E^2_\lambda\psi(\textbf{x}),
\end{equation}
$\mbox{\boldmath $\nabla$}$ is the covariant differential operator
in an external (background) field.

We are considering a static background in the
form of the cylindrically symmetric gauge flux tube of the finite
transverse size. The coordinate system is chosen in  such a way that
 the tube is along the $z$ axis.
  The tube in 3-dimensional space is obviously generalized to
 the $(d-2)$-tube in $d$-dimensional space by adding extra $d-3$
 dimensions as longitudinal ones.
 The covariant derivative is $\nabla_0=\partial_0$, $\mbox{\boldmath
$\nabla$}=\mbox{\boldmath $\partial$}-{\rm i} \tilde e\, {\bf V}$
with $\tilde e$ being the coupling constant of dimension
$m^{(3-d)/2}$ and the vector potential possessing only  one
the nonvanishing component is given by
\begin{equation}\label{4}
V_\varphi=\Phi/2\pi,
\end{equation}
outside the tube; here  $\Phi$ is the value of the gauge flux inside
the $(d-2)$-tube and $\varphi$ is the angle in  polar $(r,\varphi)$
coordinates on a plane that is transverse to the tube.   The
Neumann boundary condition at the  surface of the tube
$(r=r_0)$ is imposed on the scalar field:
\begin{equation}\label{5}
\left.\partial_r\psi_\lambda\right|_{r=r_0}=0,
\end{equation}
i.e.the surface of the flux tube is  a perfectly rigid boundary for the matter field.

The solution of \eqref{a12} satisfying the boundary condition \eqref{5} outside the impenetrable
tube of  radius $r_0$ takes form
\begin{equation}\label{6}
\psi_{kn{\bf p}}({\bf x})=(2\pi)^{(1-d)/2}e^{{\rm i}\bf{p
x}_{d-2}}e^{{\rm i}n\varphi}\Omega_{|n- {\tilde e}
\Phi/2\pi|}(kr,kr_0),
\end{equation}
where
\begin{equation}\label{7}
\Omega_\rho(u,v)=\frac{Y^\prime_{\rho}(v)J_{\rho}(u)-J^\prime_{\rho}(v)Y_{\rho}(u)}{\left[{J^{\prime}_{\rho}}^2(v)+{Y^{\prime }_{\rho}}^2(v)\right]^{1/2}},
\end{equation}
and $0<k<\infty$, $-\infty<p^j<\infty$ ($j=\overline{1,d-2}$), $n\in
\mathbb{Z}$ ($\mathbb{Z}$ is the set of integer numbers),
 $J_\rho(u)$ and $Y_\rho(u)$ are the Bessel functions of order $\rho$ of the first and  second
 kinds, the prime near the function means derivative with respect to the function argument. Solutions \eqref{6} obey orthonormalization condition
\begin{equation}\label{8}
\int\limits_{r>r_0} d^{\,d}{\bf x}\, \psi_{kn{\bf p}}^*({\bf
x})\psi_{k'n'{\bf p}'}({\bf
x})= \frac{\delta(k-k')}{k}\,\delta_{n,n'}\,\delta^{d-2}(\bf{p}-\bf{p}').
\end{equation}

The standard definition for vacuum energy density is the vacuum expectation value
of the time–time component of the energy–momentum tensor
\begin{equation}\label{a14}
\varepsilon=\langle {\rm
vac}|\left(\partial_0\Psi^+\partial_0\Psi+\partial_0\Psi\partial_0\Psi^+\right)|{\rm
vac}\rangle =\sum\hspace{-1.4em}\int\limits_{\lambda}E_\lambda\psi^*_\lambda(\textbf{x})\,\psi_\lambda(\textbf{x}).
\end{equation}
This relation suffers from ultraviolet
divergencies. The well-defined quantity  is obtained with help of regularization and then renormalization procedures, see, e.g., \cite{Most}.

For regularization, one can use zeta-function
method, see, e.g., \cite{Eli,Dow,Haw}, i.e. by inserting inverse
energy in a sufficiently high power
\begin{equation}\label{a16}
   \varepsilon_{reg}(s)=\sum\hspace{-1.4em}\int\limits_{\lambda}
   E_\lambda^{-2s}\psi^*_\lambda(\textbf{x})\,\psi_\lambda(\textbf{x}).
\end{equation}
Sums (integrals) are convergent in the case of Re\,$s > d/2$. Thus,
summation (integration) is performed in this case, and then the result will be analytically continued to the case of $s = -1/2$.

In our case the magnetic field configuration in the excluded region, irrespective of the
number of spatial dimensions, the renormalization procedure is reduced to making one subtraction, namely to subtract the contribution
corresponding to the absence of the magnetic flux, see \cite{BabSit}.

 Now, to compute  the vacuum expectation value of the
 energy density  we have to substitute \eqref{6} into
 \eqref{a16} and then obtain
 \begin{equation}\label{a29}
  \varepsilon_{ren}(s)=
 (2\pi)^{1-d}\lim_{s\rightarrow-1/2}\int d^{d-2}p\int\limits_0^\infty
  dk\,k \left({\textbf{p}}^2\!+\!k^2\!+\!m^2\right)^{-s}  [S(kr,kr_0,\Phi)\!-\!S(kr,kr_0,0)],
\end{equation}
where
\begin{equation}\label{a29a}
S(kr,kr_0,\Phi)=\sum_{n\in\mathbb
 Z}\Omega^2_{|n- {\tilde e}
\Phi/2\pi|}(kr,kr_0).
\end{equation}
Because of the infinite range of summation  $S$-function
will depend only on the fractional part of the flux
\begin{equation}\label{a29a1}
   F=\frac{\tilde e\Phi}{2\pi}-\left[\!\left[\frac{\tilde e\Phi}{2\pi}\right]\!\right],\quad(0\leq F < 1),
\end{equation}
where $[[u]]$ is the integer part of quantity u (i.e. the integer
which is less than or equal to u). So, we get
\begin{equation}\label{a29a}
S(kr,kr_0,F)= \sum_{n=0}^\infty[\Omega^2_{n+F}(kr,kr_0)+\Omega^2_{n+1-F}(kr,kr_0)]
\end{equation}
and conclude that induced vacuum energy density \eqref{a29} depends  on $F$, i.e. it is periodic in the flux $\Phi$ with a
period equal to $2\pi {\tilde e}^{-1}$. Moreover, the value of the induced vacuum energy density is symmetric under the substitution $F \rightarrow 1-F$.

In the absence of the magnetic flux in the tube $S$-function
takes the form
\begin{equation}\label{c1}
S(kr,kr_0,0)=\Omega^2_{0}(kr,kr_0)+ 2\sum_{n=1}^\infty\Omega^2_{n}(kr,kr_0).
\end{equation}

Unfortunately, the computation of  vacuum energy density in the case of the finite-thickness magnetic tube can not be done analytically because of the complicated form of $\psi$-function \eqref{6} and require numerical methods.

\section{Numerical evaluation of energy density}

In this paper, we will consider the simplest situation of $(2+1)$-dimensional space-time
and consider induced vacuum energy density outside  the impenetrable tube (it is a ring in 2-dimensional space) of radius $r_0$ with Neumann boundary condition at its edge and with half-integer values of magnetic flux $F=1/2$ inside the tube.
At this flux value, we expect
the maximal effect of vacuum polarization by analogy with the case of singular magnetic
vortex, see, e.g., \cite{Sit,our2}. Based on the results of \cite{our3} -- \cite{our2013} for computation of the  induced vacuum energy density outside the magnetic impenetrable tube with Dirichlet boundary condition at its edge we can conclude that we can immediately take $s=-1/2$ in \eqref{a29}. 

Let us now briefly discuss the main ideas of numerical calculations.
Expression \eqref{a29} is finite and can be evaluated numerically because of the possibility to restrict  the upper limit of integration and summation in it. 
To make numerical computations in this case
instead of  \eqref{a29}, it is better to use the following relation for dimensionless quantity
 \begin{equation}\label{a29Num}
r^3  \varepsilon_{ren}=
 \frac{1}{2\pi}\int\limits_0^{z_{max}}
  dz z\,  \sqrt{z^2+\left(\frac{mr_0}\lambda\right)^2}  \sum_{n=0}^{n_{max}(z)} [2\Omega^2_{n+1/2}(z,\lambda z)-\Omega^2_{n}(z,\lambda z)-\Omega^2_{n+1}(z,\lambda z)],
\end{equation}
where we introduced a dimensionless
variables
\begin{equation}\label{c5}
 kr=z,\quad\lambda=r_0/r,\quad \lambda\in[0,1].
\end{equation}
The case of $\lambda=1$ corresponds to
$r=r_0$, i.e. the point on the boundary of the tube, the case of
$\lambda=0$ corresponds to the point on the infinity
$r\rightarrow\infty$ or the case of singular tube $(r_0=0)$. 

The necessary number of terms for summation ($n_{max}(z)$) is defined at the fixed value of parameter $z$ from the condition 
that the summation result with a high precision  did not change with an increase in the number of terms. 

For small values of $z$, we make a direct integration of
function in \eqref{a29Num}. 
For large values of $z$, we use another approach.
The integrand function  in this case is a quasi-periodic\footnote{Value of period slowly decreases under increasing of the function argument.} oscillating function (with the change of sign) with  slowly decreasing amplitude with increasing 
of its argument. So, it is convenient to integrate over these periods separately with  the right value of the parameter $n_{max}(z)$. In such a way we get a falling series, each element of which is the value of the integral over the single period of the function. Getting a sufficiently large number of the elements of the series we stop integrating and interpolate the series forward. The final result of integration is the sum of the integral for small values of $z$, the sum of the explicitly counted elements of the series, and the sum of interpolated series. If the contribution to the overall integration result from the sum of the interpolated series is a few percent or less, then the computation result can be considered reliable. In the next step, we compute $r^3  \varepsilon_{ren}$ \eqref{a29Num} for different  values of parameter $\lambda$ and interpolate obtained results.

It should be noted that in the case of a singular magnetic vortex  the analytical expressions for the induced vacuum energy density can be 
obtained \cite{Sit1,our2}. For the case $(2+1)$-dimensional space-time and half-integer magnetic flux value $F=1/2$ it is expressed in terms of the Macdonald function $K_\rho(u)$ and modified Struve function $L_\rho(u)$
\begin{multline}\label{d4}
r^3 \varepsilon^{sing}_{ren}=
\frac{x^3}{3\pi^2}\left\{ \frac \pi2+
\frac{
K_0(2x)}{2x}-\left(1-\frac1{2x^2}\right)K_1(2x)- \right.\\
\left.\vphantom{\frac{K}2}- \pi x
\left[K_0(2x)L_{-1}(2x)+K_1(2x)L_0(2x)\right]\right\},\quad x=mr.
\end{multline}

The result of our computation for the induced vacuum energy density outside the impenetrable magnetic tube with Neumann boundary condition at its edge is presented in Fig.1 as a function of the dimensionless distance from the center of the tube $(mr)$ for the different values of the dimensionless tube radius $(mr_0)$. For comparison, we demonstrate also induced vacuum energy density for the case of the singular magnetic vortex.

\begin{figure}
    \centering
    \includegraphics[width=11cm]{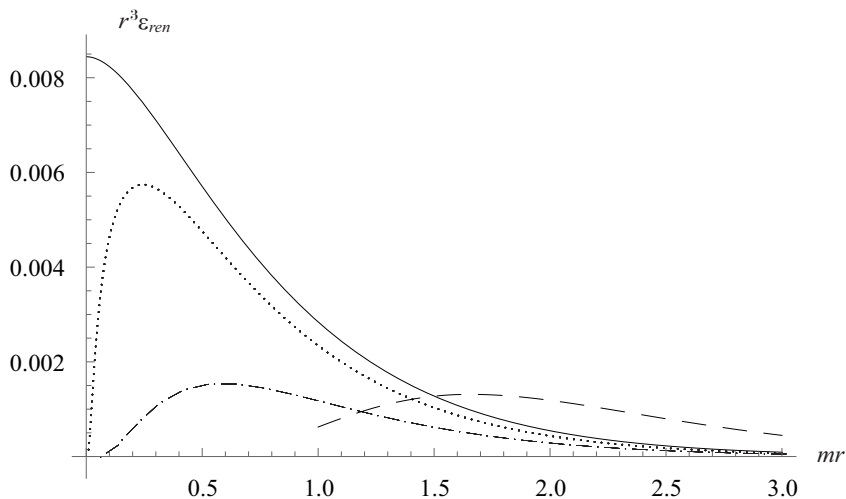}
\caption{Induced vacuum energy density of the charged scalar matter outside impenetrable magnetic
tube with Neumann boundary condition at its edge for $(2+1)$-dimensional space time and half-integer magnetic flux value $F=1/2$.
Dotted line corresponds to the case of the dimensionless tube radius  $mr_0=0.01$, dashed-dotted line to the case of $mr_0=0.1$ and dashed line corresponds to the induced vacuum energy density multiplied by 200, and $mr_0=1$. Solid line corresponds to the case of the singular magnetic vortex.}
    \label{fig:my_label}
\end{figure}

It is interesting also to compare obtained induced vacuum energy density with the case of vacuum polarization outside an impenetrable magnetic tube with perfectly reflecting (Dirichlet)  boundary condition at its edge. The results of the comparison are presented in Fig.2.

\section{Summary}

\begin{figure}[t]
    \centering
    \includegraphics[width=\textwidth]{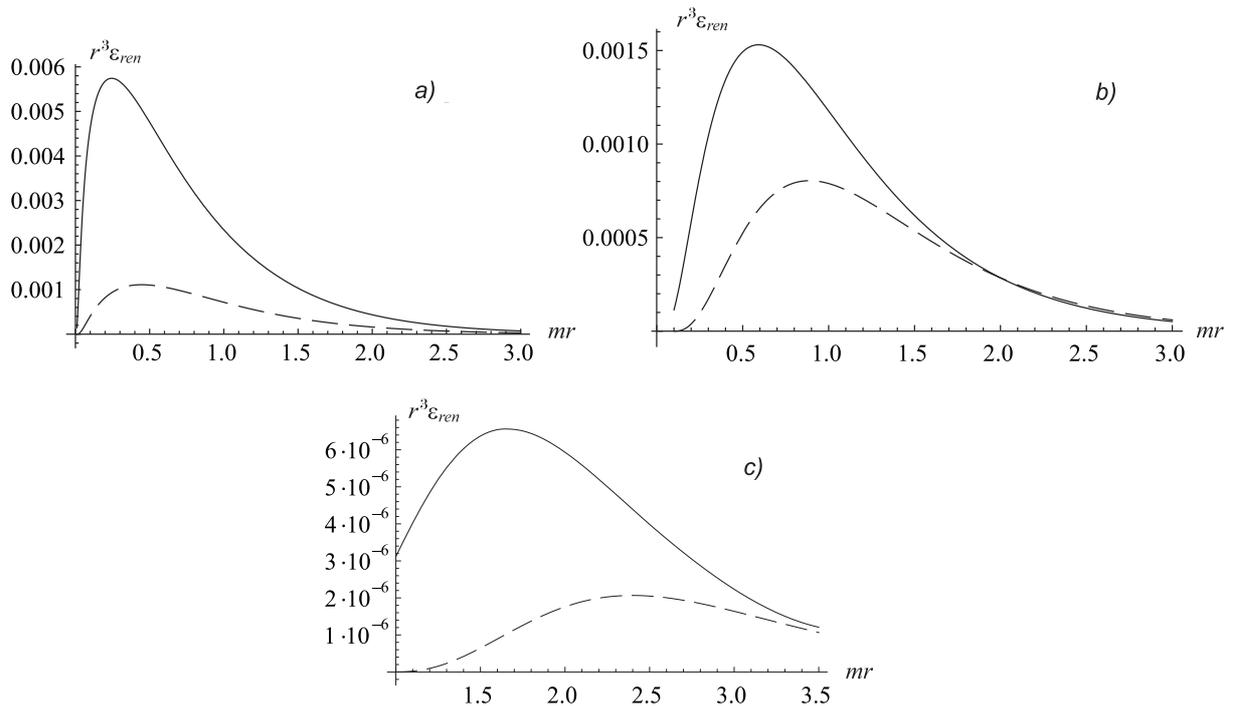}
    \caption{Comparison of the induced vacuum energy density of the charged scalar matter outside impenetrable magnetic
tube with Neumann (solid line) and Dirichlet (dashed line)  boundary conditions at its edge for the case of $(2+1)$-dimensional space time and half-integer magnetic flux value $F=1/2$: a) $mr_0=0.01$, b) $mr_0=0.1$ and dashed line corresponds to the induced vacuum energy density multiplied by 10, c) $mr_0=1$ and dashed line corresponds to the induced vacuum energy density multiplied by 1000.}
    \label{fig:my_label}
\end{figure}

We obtained general relation for computation of  the vacuum polarization of the quantized charged scalar field in the background of
a  $(d-1)$-tube (it is infinitely long tube for $d=3$ and ring for $d=2$ case) with static
magnetic field inside in flat $(d+1)$-dimensional space-time for the case when
the tube  is impenetrable for scalar
field and obeys perfectly rigid (Neumann) boundary conditions at its surface.
We showed that induced vacuum energy, in this case,
\eqref{a29a} depends periodically on the magnetic  
flux  inside the tube  with  a period equal to $2\pi {\tilde e}^{-1}$. The effect of vacuum polarization disappears for an integer value of the magnetic flux $\Phi=2\pi n {\tilde e}^{-1} $, $n\in
\mathbb{Z}$.  Thus the induced vacuum energy depends only on the fractional part of the magnetic  flux. We can see the manifestation of the  Casimir-Bohm-Aharonov effect \cite{Sit} in this case. 

Our results confirm the statement that the  Casimir-Bohm-Aharonov effect 
is due to the condition of the impenetrability  of the tube for the matter field. Otherwise, namely in the case when quantized matter penetrates into the region with a magnetic field,  the dependence of  the  induced vacuum polarization effect from  the magnetic flux is not periodic: the effect is determined by the value of the total magnetic flux in the tube, see, e.g., \cite{Cangemi} -- \cite{Graham}.

In the simplest case of $(2+1)$-dimensional space-time, with help of numerical methods, we compute the value of the vacuum energy density of the quantized charged scalar field outside  the impenetrable tube of radius $r_0$ with Neumann
boundary conditions at its edge. We chose half-integer value of the magnetic flux  $F=1/2$ inside the tube.
At this flux value, we expect the maximal effect of vacuum polarization by analogy with the case of singular magnetic
vortex, see, e.g., \cite{Sit,our2}.  We made computations  without the regularization procedure of the vacuum energy but due to its renormalization by subtracting the contribution corresponding to the absence of the
magnetic flux, see \eqref{a29Num}.

The results of our computations are presented in Fig.1.
One can see, that at the same dimensionless distance from the center of the tube ($mr$) the vacuum polarization effect is largest for the case of the singular magnetic vortex and exponentially quickly decrease with the growth of the tube radius. It should be noted, that effect of vacuum polarization becomes negligible when the radius of the tube is of order or more than the Compton wavelength of the matter field ($mr_0\gtrsim 1$).

The comparison of the vacuum polarization in the case of the perfectly rigid (Neumann) and the perfectly reflecting (Dirichlet) boundary conditions at the tube edge is presented in Fig.2. One can see, that for the tubes of the same thickness the vacuum polarization effect is always the largest in the case of the Neumann boundary condition. This result is in agreement with the result of \cite{our2022} for the case of induced magnetic flux.

We need to pay attention to that convergence of the integral under the computation of the induced vacuum energy density \eqref{a29Num} in the case of the Neumann boundary condition at the tube edge is sufficiently weakly than in the case of the Dirichlet boundary condition. The complexity of the computations strongly increases with decreasing in the tube thickness. We conclude that relation \eqref{a29Num} based on the direct usage of the field solutions \eqref{6} is not suitable for the computations of the vacuum polarization outside the impenetrable thin  tube ($mr_0\ll 1$). In this case, more appropriate, in our opinion, should be the technique  of computation with help of transformation in the complex plane, when Bessel functions $J_\nu(y)$ and $Y_\nu(y)$ transform to the modified Bessel $I_\nu(y)$ and Macdonald $K_\nu(y)$ functions, see, e.g., \cite{our2022prd}.

\section*{Acknowledgments}

The work of Yu.A.S. was supported by the National Academy of Sciences of Ukraine (Project No. 0122U000886).


\end{document}